\documentclass[aps,pra,floatfix,showpacs,twocolumn,superscriptaddress]{revtex4}
\usepackage{epsfig}
\usepackage{subfigure}

\newcommand{\ket}[1]{\mbox{$|#1\rangle$}}

\newcommand{\braket}[2]{\mbox{$\langle #1|#2\rangle$}}
\newcommand{\op}[1]{\mbox{\boldmath $\hat{#1}$}}

\begin{document}

\draft

\title{Entanglement Creation Using Quantum Interrogation}

\author{A.~Gilchrist}\email{alexei@physics.uq.edu.au}
\affiliation{Department of Physics, University of Queensland, QLD 4072, Brisbane, Australia.}
\author{A.~G.~White}
\affiliation{Department of Physics, University of Queensland, QLD 4072, Brisbane, Australia.}
\author{W.~J.~Munro}
\affiliation{Hewlett-Packard Laboratories, Filton Road, Stoke Gifford, Bristol, BS34 8QZ, United Kingdom.}

\date{\today}

\begin{abstract}
  We present some applications of high efficiency quantum interrogation
  (``interaction free measurement'') for the creation of entangled states
  of separate atoms and of separate photons. The quantum interrogation of a  quantum object in a superposition of object-in and object-out leaves the
  object and probe in an entangled state. The probe can then be further
  entangled with other objects in subsequent quantum interrogations. By
  then projecting out those cases were the probe is left in a particular
  final state, the quantum objects can themselves be left in various
  entangled states. In this way we show how to generate two-, three-, and
  higher qubit entanglement between atoms and between photons.  The effect
  of finite efficiency for the quantum interrogation is delineated for the
  various schemes.
\end{abstract}


\maketitle

\section{Introduction}

Quantum Information processing is currently receiving considerable
attention \cite{00ncQCQI,bill-2}, with significant effort focussed on
finding applications. Known applications include quantum
computation \cite{95v255,98vp1}, quantum communication \cite{96s2614},
quantum cryptography \cite{91e661,00tbzg4737,00jswwz4729,00npwbk4733},
quantum teleportation \cite{98bbmhp1121,93bbcjpw1895,97bpmewz575}, quantum
dense coding \cite{00bk042302} and high precision
measurements \cite{97hmpepc3865,0109049}.  At the heart of many of these
applications is entanglement, which is generally thought to be one of the
key resources required in quantum information processing. The
characterisation of entangled states and entanglement is a
challengingproblem and a considerable theoretical effort has been invested
in characterising entanglement in a variety of physical situat ions
\cite{0109049,96bdsw3824,98w2245,97vprk2275,98vp1619,99jp3566,99hhh1056,97bvpkh3327,96p1413,00dmn225}.
Likewise there has been considerable experimental effort i n developing
techniques for creating highly entangled resources (e.g entangled photons
\cite{photon} or ions \cite{ion}), including the ability to produce
arbitrary entangled states \cite{white99,white01}.

In this paper we propose several schemes using quantum interrogation (QI)
to generate entanglement between the states of separate particles,
expanding on a suggestion in reference~\cite{96kwz673}.  The
technique of quantum interrogation (also known as ``interaction free
measurement'') has its roots in ``negative results'' measurements
originally discussed by Renninger \cite{60r417}, and later by Dicke
\cite{81d925} who analyzed the change in an atom's wave-function by the
\emph{non-scattering} of a photon from it. In 1993 Elitzur and Vaidman (EV)
proposed a particularly dramatic version where a photon was used to
ascertain the presence of a light sensitive bomb without the bomb
exploding, hence seemingly without interacting with it \cite{93ev987}.  The
EV scheme works with at best an efficiency of 50\%, i.e. at most 50\% of
the measurements are ``interaction free''.  High efficiency schemes making
use of the quantum Zeno effect \cite{77ms756} were proposed by Kwiat
\emph{et al.} \cite{99kwmnwwz4725} and achieved an efficiency of 74\%. An
alternative scheme using high finesse resonators was introduced by
\cite{98tgkbll3987} and achieved a comparable efficiency.  The above
figures take into account other losses that we will not consider, so to
avoid confusion we will characterize out figures of merit against the
number of cycles in a QI.

\begin{figure}
  \begin{center}
    \subfigure[]{\epsfig{figure=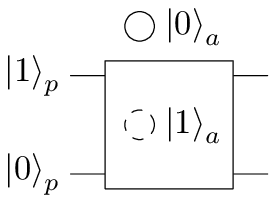}}\quad
    \subfigure[]{\epsfig{figure=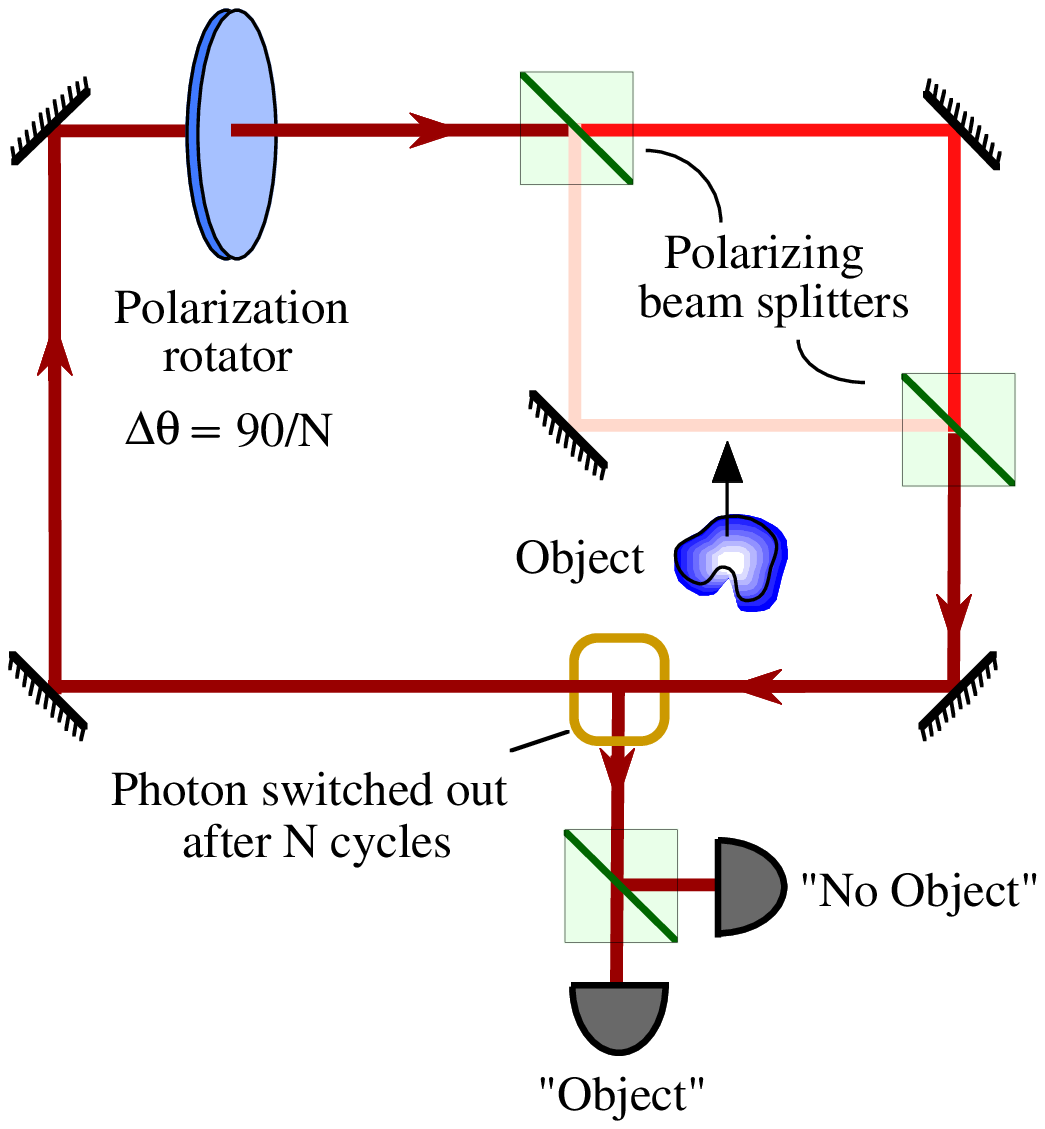,width=.5\columnwidth}}
  \end{center}
  \caption{ (a) An idealized quantum interrogation and the labeling of logical 
    qubits. $\ket{0}_p$ and $\ket{1}_p$ are the logical states of the probe
    particle and $\ket{0}_a$ and $\ket{1}_a$ are the logical states of the
    object. (b) An optical implementation of high-efficiency quantum
    interrogation.  The probe particle is a photon for which
    horizontal and vertical polarization represent the target qubit state
    and the presence or absence of an absorbing object represents the
    control qubit state.}
  \label{fig:qi_ideal}
\end{figure}

Consider an idealized high-efficiency quantum interrogation scheme, of the
type presented in \cite{99kwmnwwz4725}, in the limit of perfect efficiency.
We shall take the absorbing object to be a quantum device that can be in
one of two states: $\ket{0}_a$ representing object-out, i.e. a completely
transparent object; and $\ket{1}_a$ representing object-in, i.e. a
completely absorbing object.  We shall probe the state of the object using
a photon which can be in one of the two states $\ket{0}_p$ or $\ket{1}_p$
which can be represented schematically as two ports to the device as in
figure~\ref{fig:qi_ideal}(a). The two states of the photon could be, for
example, different polarization states as in figure~\ref{fig:qi_ideal}(b)
(figure taken from Kwiat \emph{et al.}  \cite{99kwmnwwz4725}) or different
spatial modes.

Quantum interrogation functions in the following way: with the object out a
probing photon initially in state $\ket{0}_p$ or $\ket{1}_p$ remains
unchanged and exits the device in the same state [as in
figure~\ref{fig:qi_ideal}(b) with the addition of a $90\deg$ polarization
rotation at the end]. With the absorbing object in state $\ket{1}_a$
(object-in) then a photon initially in state $\ket{0}_p$ will evolve to
state $\ket{1}_p$ without changing the state of the object (an
``interaction free measurement''). If we probe the object with a photon
initially in state $\ket{1}_p$ the photon will certainly get absorbed by
the object --- this event was dramatized as a bomb exploding in the EV
scheme.

With this representation the behavior of the quantum
interrogation is tantalizingly close to the operation of a CNOT gate.  That
is we have the mapping $Q$:
\begin{equation}
\label{eq:qi_logic}
Q:\begin{array}{rcl}
  \ket{00}&\rightarrow&\ket{00}\\
  \ket{01}&\rightarrow&\ket{01}\\
  \ket{10}&\rightarrow&\ket{11}\\
  \ket{11}&\rightarrow&\ket{\mbox{\it boom!\/}}
  \end{array}
\end{equation}
where the first mode represents the state of the object and the second the
state of the photon. We could equally have flipped the interpretation of
the two ports so that with the object in we would have
$\ket{11}\rightarrow\ket{10}$ and $\ket{10}\rightarrow\ket{\mbox{\it
    boom!\/}}$, we shall represent this alternative map as $Q_r$.  It
should be noted that since only a single combination of the terms in the
map~(\ref{eq:qi_logic}) fails, if we can detect the failure event
(detecting the bomb exploding) then we could in principle recreate the
appropriate state.  We shall, however, assume that this is not possible for
the purposes of this paper.

\begin{figure}
  \begin{center}
    \epsfig{figure=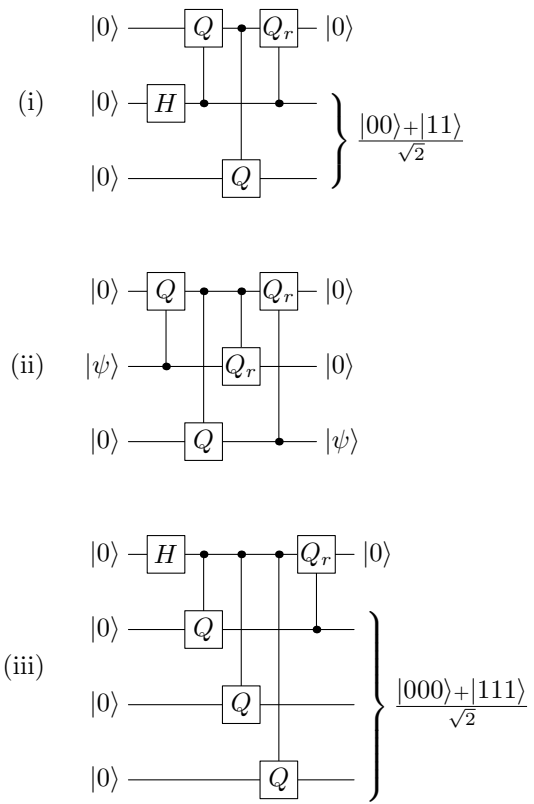}
\end{center}
  \caption{Several quantum circuits constructed using a perfect efficiency
    quantum interrogation measurement, $Q$, and, $Q_r$, which are explained
    in the text. $H$ is a Hadamard gate.  Circuit (i) creates  Bell
    states. Circuit (ii) is a quantum bus, which swaps a qubit from one
    channel to another. Circuit (iii) creates a GHZ state.  }
    \label{fig:ideal_circuits}
\end{figure}

Despite not having access to the full logic table for
a CNOT, the device proves remarkably useful as can be seen from some of the
quantum circuits that can be constructed using it depicted in
figure~\ref{fig:ideal_circuits}.
There are three principal obstructions to performing these ideal circuits:
\begin{enumerate}
\item The effect of finite efficiency in the quantum interrogation scheme.
\item The potential inability to switch the r\^oles of the control and
  target. For instance, it is much easier to have an interferometer using
  photons (the target) and a suitable atom as the quantum object (the
  control), than to have an atom interferometer repeatedly probing the
  state of a single photon.
\item The effect of a semi-transparent object, for instance see
  reference~\cite{01mm032105,99wkj268}.

\end{enumerate}

In this paper we shall examine the first two issues and leave the third for
a subsequent work.  In the schemes that follow we shall restrict ourselves
to using the state of some atom as the control qubit and the state of a
photon as the target qubit. In section~\ref{sec:model} we present a simple
model of a quantum interrogation measurement of a specific quantum object.
In section~\ref{sec:cond-entangl-prep} we propose three conditional schemes
to generate Bell, $W$, and GHZ type entanglement in the state of two and
three atoms using photons as mediators. In
section~\ref{sec:determ-entangl-prep} we propose using an atom to generate
Bell and GHZ type entanglement between separate photons.

\section{The Model}
\label{sec:model}

\begin{figure}
  \begin{center}
    \epsfig{figure=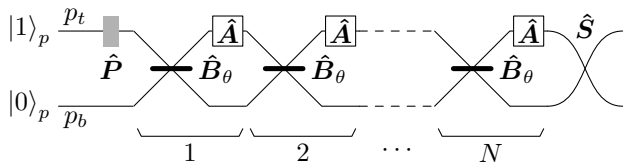}
\end{center}
  \caption{Schematic representation of a high-efficiency quantum 
    interrogation. A single photon probes the state of an atom $\op{A}$,
    through repeated passes through a Mach-Zender interferometer. The flow
    of time is to the right. Note that there is an initial $180^\circ$
    phase shift $\op{P}$ applied to the top arm, and a final interchange of
    the modes in order to achieve a more convenient logic structure.}
  \label{fig:qi_flattened.eps}
\end{figure}

We can represent the quantum interrogation apparatus as a series of $N$
Mach-Zender interferometers laid end on end as in
figure~\ref{fig:qi_flattened.eps} where it is understood that the absorbing
object labelled $\op{A}$ in the figure is the same object each time. This
is equivalent to the experimental arrangement in
figure~\ref{fig:qi_ideal}(b).  We shall label the light modes above and
below the beam splitters as modes $p_t$ and $p_b$ respectively.  Thus a
photon in the top mode ($\ket{1}_{p_t}\ket{0}_{p_b}$) will be used to code
a logical $\ket{1}_p$ for the photon qubit, and a photon in the bottom mode
($\ket{0}_{p_t}\ket{1}_{p_b}$) will code a logical $\ket{0}_p$ for the
photon qubit.

\begin{figure}
  \begin{center}
    \epsfig{figure=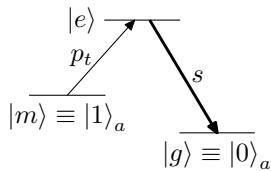}
  \end{center}
  \caption{Model of the interaction with the atom and the labeling 
    of the logical basis. The levels $\ket{m}$, $\ket{e}$ and $\ket{g}$ are
    meta-stable, excited and ground states respectively. A photon in
    the top mode of the quantum interrogation ($p_t$) can induce a coherent
    evolution between states $\ket{m}$ and $\ket{e}$. State $\ket{e}$
    experiences rapid decay to the ground state, releasing a scattered
    photon $s$. 
  }
  \label{fig:atom_levels.eps}
\end{figure}

We shall take as our model of the absorbing object a three level atom,
similar to that introduced in \cite{00pssrmm060101}, depicted in
figure~\ref{fig:atom_levels.eps}. The atom can start in a meta-stable state
$\ket{m}$, from which it will absorb a photon from mode $p_t$ with unit
efficiency. After absorbing a photon the atom immediately decays from the
excited state $\ket{e}$ to the ground state $\ket{g}$ which is far
off-resonance from the meta-stable state. We are able to neglect the
re-absorption of the emitted photon so this forms an essentially
irreversible process. We can then label the meta-stable state as our logical
$\ket{1}_a$ (object-in) for the atom qubit.  The atom in its ground state
is transparent to the $p_t$ photons, and so we can label the ground state
as our logical $\ket{0}_a$ (object-out) for the atom qubit. Note that
filtering off the higher frequency scattered photons removes the problems
of forward scattering \cite{99wkj268}.

In what follows the atom is always considered to be the control qubit and
the photon the target qubit and we shall always write them in that order.
We shall use the subscripts $p$ and $a$ to denote photon and atom only if
necessary.

The effects of the atom and beam splitters on the modes (in the logical
basis) are then:
\begin{eqnarray}
  \label{eq:transformations}
  \op{A}:&&
    \begin{array}{rcl}
        \ket{1}_a\ket{0}_p &\rightarrow& \ket{1}_a\ket{0}_p\\
        \ket{1}_a\ket{1}_p &\rightarrow& \ket{0}_a\ket{s}_p
    \end{array}\\
  \op{B}_\theta:&&
  \begin{array}{rcl}
    \ket{0}_p&\rightarrow&\cos\theta\ket{0}_p+\sin\theta\ket{1}_p\\
    \ket{1}_p&\rightarrow&\cos\theta\ket{1}_p-\sin\theta\ket{0}_p
  \end{array}
\end{eqnarray}
where the reflectivity $R=\cos^2\theta$, and $\ket{s}_p$ represents a
scattered photon. Note that a photon being absorbed and scattered by the
atom removes the system from the logical basis (there will be no photon in
either $p_t$ or $p_b$) and in writing the state $\ket{s}$ we are using a
convenient shorthand to denote this event.

After $N$ cycles within the quantum interrogation, with the atom and photon
initially in state $\ket{\phi_0}$, we will evolve to the state,
\begin{equation}
  \label{eq:state_N}
  \ket{\phi_N}=\op{S}\op{A}_N\op{B}\op{A}_{N-1}\ldots\op{B}\op{P}\ket{\phi_0},
\end{equation}
where $\op{P}$ is a $180^\circ$ phase shift and $\op{S}$ represents the
final swap of the modes --- these operations are done to achieve a ``nice''
logic structure.

With the atom in state $\ket{0}_a$ (object-out) after $N$-cycles we have:
\begin{eqnarray}
  \label{eq:0a}
  \ket{0}_p&\rightarrow& \cos(N\theta)\ket{1}_p+\sin(N\theta)\ket{0}_p\\
  \ket{1}_p&\rightarrow& -\cos(N\theta)\ket{0}_p+\sin(N\theta)\ket{1}_p
\end{eqnarray}
We choose
$\theta=\pi/2N$ so that $\ket{0}_p\rightarrow\ket{0}_p$ and similarly
$\ket{1}_p\rightarrow\ket{1}_p$.

Now consider the atom initially in the state $\ket{1}_a$ (object-in), After
$N$ cycles equation~(\ref{eq:state_N}) yields
\begin{eqnarray}
  \label{eq:10}
  \ket{1}_a\ket{0}_p &\rightarrow&\cos^N\theta\ket{11}
+\sin\theta\sum_{j=0}^{N-1}\cos^j\theta\ket{0s_j}\\
  \label{eq:11}
\ket{1}_a\ket{1}_p &\rightarrow&
\sin\theta\cos^{N-1}\theta\ket{11}\nonumber\\
&&-\cos\theta\ket{0s'}+ \sin^2\theta\sum_{j=0}^{N-2}\cos^j\theta\ket{0s_j}
\end{eqnarray}
where we have dropped the subscripts for the kets on the right.

In the limit that $N\rightarrow\infty$ then all the terms with a
$\sin\theta$ disappear and equations~(\ref{eq:0a})--(\ref{eq:11}) show the
logic given in~(\ref{eq:qi_logic}).

\section{Atom Entanglement Preparation}
\label{sec:cond-entangl-prep}

In this section we present schemes for generating several types of
entangled states between atoms of the type described in
section~\ref{sec:model} using photons as a mediating particle. These
schemes allow the entanglement of separated atoms without ever bringing
them into direct interaction with each other. All the schemes are
non-deterministic in that they will work only a certain percentage of the
time, when a specific result is obtained upon measuring the photon. This is
a limitation that is common to many entanglement generation schemes.
There is an added advantage in
using a conditioned scheme in our case. Detecting the final state of the
photon in either $\ket{0}_p$ (photon in mode $p_b$) or in $\ket{1}_p$
(photon in mode $p_t$) means we condition out those cases where the atom
absorbs a photon since the photon will be removed from both modes of the
interferometer.  This guarantees that we generate a pure, entangled, state.

To characterize the success of a scheme in generating a particular
entangled state we shall use the Fidelity\footnote{It is also possible to
  define the Fidelity as $F=|\braket{\psi_{\mbox{\small
        desired}}}{\psi_{\mbox{\small actual}}}|^2$, and both definitions
  are used in the literature. Here we've choosen to follow the convention
  in \cite{00ncQCQI}} $F$ which is simply
\begin{equation}
  \label{eq:fidelity}
  F=|\braket{\psi_{\mbox{\small desired}}}{\psi_{\mbox{\small actual}}}|
\end{equation}
and the tangle $\tau$, which is the square of the concurrence \cite{98w2245}
from which the entanglement of formation can be calculated.
For a mixed state $\rho$ of two qubits, the concurrence $C$ if given by
\begin{equation}
  \label{eq:concurrence}
  C=\mbox{max}(\lambda_1-\lambda_2-\lambda_3-\lambda_4,0)
\end{equation}
where the $\lambda_i$ are the square roots of the eigenvalues, in
decreasing order, of $\rho\tilde\rho = \rho\,\sigma_y^A\otimes\sigma_y^B
\rho^* \sigma_y^A\otimes\sigma_y^B$ and $\rho^*$ denotes the complex
conjugation of $\rho$ in the computational basis
$\{\ket{00},\ket{01},\ket{10},\ket{11}\}$.

The tangle is valid for two qubits; for three qubits in a pure state we
will use the 3-tangle \cite{00ckw052306} which gives the purely three way
entanglement of the system:
\begin{equation}
  \label{eq:3tangle}
  \tau_3=\tau_{A(BC)}-\tau_{AB}-\tau_{AC}
\end{equation}
and can be understood loosely to embody the amount of entanglement of qubit
$A$ with qubits $B$ and $C$ over and above the amount of entanglement of
qubit $A$ with $B$ and of $A$ with $C$.

% Nondet Bell ------------------------------------------------------------

\begin{figure}
  \begin{center}
    \epsfig{figure=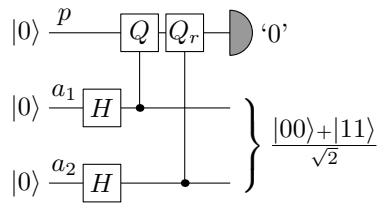}
\end{center}
  \caption{Non-deterministic generation of the Bell state 
    $(\ket{00}+\ket{11})/\sqrt{2}$. Two atoms are initially prepared in
    superposition states by Hadamard transformations. A photon is then used
    to probe each atom in turn using quantum interrogation. In the
    sub-ensemble of cases were the final state of the photon is measured
    to be $\ket{0}_p$ (mode $p_t$) the atoms have been left in the required
    Bell state. }
  \label{fig:nondet_bell}
\end{figure}

Consider the scheme depicted in figure~\ref{fig:nondet_bell}. Two atoms are
initially placed into in a superposition state. A photon makes a QI of the
first atom, and is then used to make another QI of the
second atom, where upon it's measured in the state $\ket{0}_p$ (i.e.
exiting in mode $p_b$ of the last QI). In the limit of high efficiency QI,
the two atoms will be left in a maximally entangled Bell state.

If initially we let the atoms be in arbitrary superposition states, i.e. we
have
\begin{equation}
  \label{eq:1}
  \ket{\psi_0}=(\alpha_1\ket{0}_{a_1}+\beta_1\ket{1}_{a_1})
(\alpha_2\ket{0}_{a_2}+\beta_2\ket{1}_{a_2})\ket{0}_p
\end{equation}
then after $N$ cycles within each QI, the
final state of the system \emph{conditioned} on a successful measurement
of the state $\ket{0}_p$ is
\begin{eqnarray}
  \ket{\psi_N}&=&{\mathcal{N}}
  \{\alpha_1\alpha_2\ket{00}
+\beta_1\beta_2c^{2N}\ket{11} \nonumber \\
&&+sc^{N-1}\alpha_1\beta_2\ket{01}\}
  \label{eq:nondetbellfinal}
\end{eqnarray}
where $c=\cos\theta$, $s=\sin\theta$; and the normalization ${\mathcal{N}}$
is determined by the requirement that $\braket{\psi_N}{\psi_N}=1$ after the
state is conditioned on a successful measurement.

\begin{figure}
  \begin{center}
    \epsfig{figure=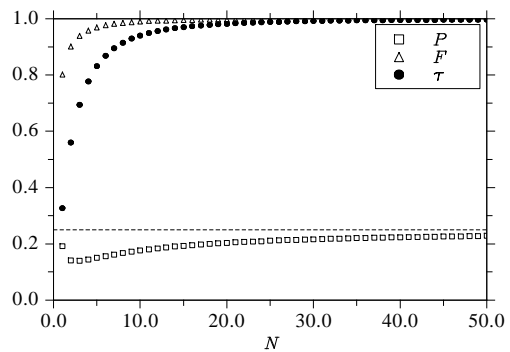,width=\columnwidth}
\end{center}
  \caption{The conditional generation of a Bell state, following the scheme in 
    figure~\ref{fig:nondet_bell}. Plotted as a function of the number
    cycles $N$ through each QI is (i) the probability of
    successful operation $P$, which has a limiting value of $1/4$ (dashed
    line), (ii) the Fidelity against the desired Bell state $F$, and (iii)
    the tangle $\tau$ of the output state.}
  \label{fig:plot_nondet_bell}
\end{figure}

In figure~\ref{fig:plot_nondet_bell} we plot the probability of successful
operation $P$, the fidelity $F$ and the tangle $\tau$ against the number of
cycles in each QI for generating the Bell state
$(\ket{00}+\ket{11})/\sqrt{2}$, with $\alpha_{1,2}=\beta_{1,2}=1/\sqrt{2}$.

We can access the other Bell states [$(\ket{01}\pm\ket{10})/\sqrt{2}$] by
either swapping the second quantum interrogation from $Q$ to $Q_r$ and
conditioning on the detection of $\ket{1}_p$, which amounts to swapping the
ports of one of the quantum interrogations; or by using local operations on
the final state.  We can therefore tune our device to produce a desired
type of entanglement.

% Nondet W ------------------------------------------------------------

We can extend the technique to three atoms, and generate an entangled
three-qubit state. We will present two schemes to generate two types of
three-qubit entanglement, which are inequivalent under local operations and
classical communication (LOCC) \cite{00dvc062314}.

\begin{figure}
  \begin{center}
    \epsfig{figure=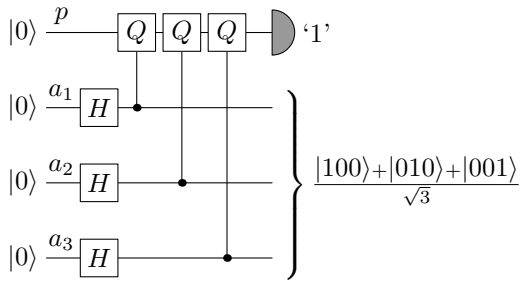}
\end{center}
  \caption{A non-deterministic preparation of the $W$-state using an 
    auxiliary mode. Atoms initially prepared in superposition states are
    probed in turn using QI. In the cases were the final
    state of the photon is $\ket{1}_p$ the atoms have been left in a $W$
    entangled state. }
  \label{fig:nondet_w_lt}
\end{figure}

Firstly we will examine the scheme in figure~\ref{fig:nondet_w_lt} for
generating the $W$ entangled state,
$\ket{W}=(\ket{001}+\ket{010}+\ket{100})/\sqrt{3}$.  With three atoms
initially in superpositions, the photon probes each atom in turn with a QI
before being detected in the state $\ket{1}_p$.

\begin{figure}
  \begin{center}
    \epsfig{figure=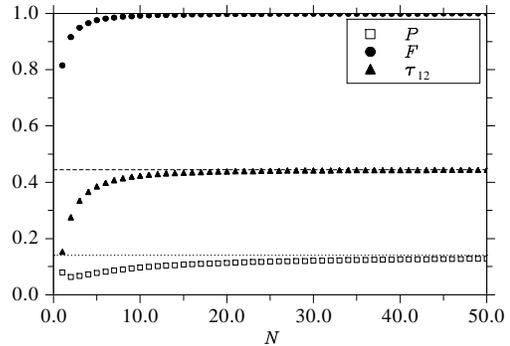,width=\columnwidth}
\end{center}
  \caption{The conditional generation of a $W$-state following the scheme in 
    figure~\ref{fig:nondet_w_lt}. Plotted as a function of the number
    cycles $N$ through each QI is (i) the probability of successful
    operation $P$, with a limiting value of $9/64$ shown as a dotted line;
    (ii) the Fidelity against the target $W$-state $F$; (iii) the tangle
    between pairs of qubits $\tau$ (all pairs have equal tangle), the
    theoretical limiting value of $4/9$ is shown as a dashed line.}
  \label{fig:plot_nondet_w}
\end{figure}

The $\ket{W}$ state has only pair-wise entanglement, so we plot the tangle
between pairs of qubits in figure~\ref{fig:plot_nondet_w}, together with
the probability of success and the fidelity. For an ideal $\ket{W}$ state the
tangle between pairs of qubits is $\tau=4/9$.

If each atom starts in an arbitrary superposition of 
$\alpha_j\ket{0}_{a_j}+\beta_j\ket{1}_{a_j}$ where $j$ indexes the atoms,
then after $N$ cycles in each QI following the scheme
in figure~\ref{fig:nondet_w_lt}, we obtain
\begin{eqnarray}
  \ket{\psi_N}&=&{\mathcal{N}}\{
c^N(\beta_1\alpha_2\alpha_3\ket{100}+
\alpha_1\beta_2\alpha_3\ket{010}+
\alpha_1\alpha_2\beta_3\ket{001})
\nonumber\\
&&sc^{2N-1}(\beta_1\beta_2\alpha_3\ket{110}
+\beta_1\alpha_2\beta_3\ket{101}
+\alpha_1\beta_2\beta_3\ket{011})\nonumber\\
&& +s^2c^{3N-2}\beta_1\beta_2\beta_3\ket{111}\}
  \label{eq:nondetwfinal}
\end{eqnarray}
In figure~\ref{fig:plot_nondet_w} are plotted various performance parameters
against $N$ for generating the $\ket{W}$ state starting with a symmetric
superposition in each atom.

As before we can access other $W$-states either by changing a $Q$ to a
$Q_r$ and conditioning on a $\ket{1}_p$, or by using local operations on
the final state. By extending the circuit in figure~\ref{fig:nondet_w_lt}
in the obvious way to more modes we can create higher order $W$-states 
such as $(\ket{1000}+\ket{0100}+\ket{0010}+\ket{0001})/2$.

% Nondet GHZ ------------------------------------------------------------

\begin{figure}
  \begin{center}
    \epsfig{figure=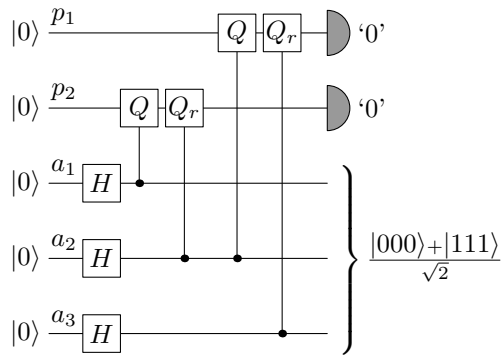}
\end{center}
  \caption{A non-deterministic preparation of the GHZ state using two 
    auxiliary modes. After three atoms have been prepared in superposition
    states, a photon probes atoms $a_1$ and $a_2$ using QI.  Another photon
    probes atoms $a_2$ and $a_3$ also using QI. In the cases were the final
    state of both photons is $\ket{00}_{p_1p_2}$ the atoms are left in the
    GHZ state shown.}
  \label{fig:nondet_ghz_lt}
\end{figure}

Finally we can use this technique to induce a GHZ state in three separated
atoms by using two auxiliary photons as depicted in
figure~\ref{fig:nondet_ghz_lt}. Here, with the atoms prepared in
superposition states, the first photon probes atoms one and two in turn,
and the second photon probes atoms two and three in turn, before both
photons are detected in the joint state $\ket{00}_{p_1p_2}$.

\begin{figure}
  \begin{center}
    \epsfig{figure=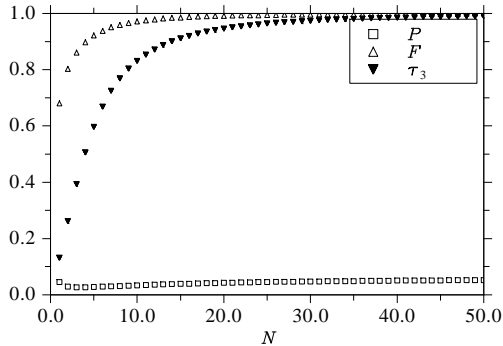,width=\columnwidth}
\end{center}
  \caption{The conditional generation of a GHZ state following the scheme in 
    figure~\ref{fig:nondet_ghz_lt}. Plotted as a function of the number
    cycles $N$ through each QI is (i) the probability of successful
    operation $P$ which has a limiting value of $2/64$, (ii) the Fidelity
    against the target Bell state $F$, and (iii) the 3-tangle $\tau_3$ of
    the output state.}
  \label{fig:plot_nondet_ghz}
\end{figure}

With the atoms each initially in the arbitrary superposition states
$\alpha_j\ket{0}_{a_j}+\beta_j\ket{1}_{a_j}$ where $j$ indexes the atoms,
then after $N$ cycles in each QI we get
\begin{eqnarray*}
    \ket{\psi_N}&=&{\mathcal{N}}\{\alpha_1\alpha_2\alpha_3\ket{000}
+c^{4N}\beta_1\beta_2\beta_3\ket{111}\nonumber\\
&&+sc^{N-1}\alpha_1\alpha_2\beta_3\ket{001}
+sc^{3N-1}\alpha_1\beta_2\beta_3\ket{011}\}
  \label{eq:nondetwfinal2}
\end{eqnarray*}
In figure~\ref{fig:plot_nondet_ghz} we characterize the success of generating 
the state $(\ket{000}+\ket{111})/\sqrt{2}$ with the three-way tangle
$\tau_3$, and the fidelity $F$, for atoms initially in equal
superposition states.

It should be noted that the circuit in figure~\ref{fig:nondet_bell}
is embedded within the circuit in figure~\ref{fig:nondet_ghz_lt} and in fact
the construction can be extended recursively to generate states
of the form $(\ket{0000}+\ket{1111})/\sqrt{2}$ and higher.
Also, as in the previous cases we can access other GHZ states.

It should be emphasized that in the three schemes presented in this
section, the post selection ensures that the final states are pure states,
as it selects specifically the cases where incoherent evolution has not
occurred.

\section{Photon Entanglement Preparation}
\label{sec:determ-entangl-prep}

\begin{figure}
  \begin{center}
    \epsfig{figure=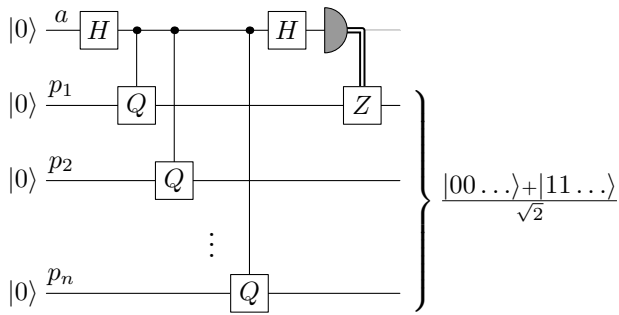}
\end{center}
  \caption{Using a measurement with a classically conditioned $Z$ gate 
    (Pauli $\sigma_z$) to replace one of the quantum interrogations.}
  \label{fig:det_entangle_lt}
\end{figure}

In the previous section we used a photon to entangle separate atoms, in
this section we'll present a scheme to use an atom to entangle independent
photons. With an atom in a superposition state we probe its state using $n$
photons, in $n$ consecutive QI's as in figure~\ref{fig:det_entangle_lt}.
Measurement of the final state of the atom can be used to
classically condition a gate (a Pauli $\sigma_z$ transformation) on one 
of the photons.

An advantage of this scheme is that in the ideal quantum interrogation
limit, it works deterministically --- it is not conditioned on the detection
of a particular result.

\begin{figure}
  \begin{center}
    \epsfig{figure=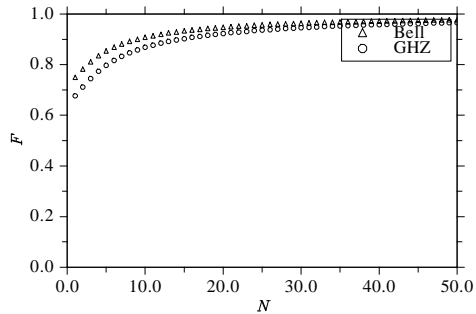,width=.9\columnwidth}
  \end{center}
  \caption{Fidelity of output state for the circuits in 
    figure~\ref{fig:det_entangle_lt} to generate $\ket{00}+\ket{11}$
     and $\ket{000}+\ket{111}$, when compared against those
    states.}
  \label{fig:plot_det_bell_ghz_f.eps}
\end{figure}

For this scheme and for finite $N$ we have a more limited group of
measures of how close we are to the ideal scheme. Where as in the previous,
atom entangling, schemes the post selection ensured the final states would
be pure, this is not the case for the photon entangling scheme.  This means
that not only will we end up with mixed states if we trace over the
environment but some of those states will be outside our logical basis (for
instance the case where there are no photons in either the top or the
bottom mode). For this reason we shall only plot the fidelity against the 
desired state (in the ideal case where there are no absorptions).
In figure~\ref{fig:plot_det_bell_ghz_f.eps} we plot the fidelity of the
output state compared with the desired state, for circuits to generate a
Bell state ($(\ket{00}+\ket{11})/\sqrt{2}$) and a GHZ state 
($(\ket{000}+\ket{111})/\sqrt{2}$).
Although the convergence is not as rapid as for the atom-entangling schemes
we still approach the desired state in relatively few cycles.

\section{Conclusions}

To conclude, in this paper we have described how high efficiency quantum
interrogation can be used to generate entangled particles.
The protocols provide a mechanism by which two or more atoms can be
entangled via a mediating photon (the photon can be thought of acting as a
bus) or alternatively how two or more photons can be entangled via a mediating
atom.

The attractive aspects of the proposal are the entanglement is created
without making use of prior entangled states; the entanglement is tunable
(i.e. using the same apparatus allows you to set the degree and type of
entanglement, including accessing different classes of higher-order
entanglement); and that for the atoms the entanglement is achieved in-situ,
without needing to bring the atoms in proximity of each other.

Although the scheme presented here is idealised (perfect optical elements
and no losses) a high degree of entanglement is achieved in remarkably few
cycles in the quantum interrogation, leading to a hope that in a real
applications, entanglement by these schemes may be achievable with current
technology.

\section*{Acknowledgements}

This research was supported by the New Zealand Foundation for Research,
Science and Technology under grant UQSL0001, AG would like to thank G.J.
Milburn for stimulating and helpful discussions. AGW wishes to acknowledge
D.F.V. James for previous discussions.

%=====================================================================
\bibliographystyle{prsty}

%\bibliography{/home/alexei/physics/ref/ref_all.bib,qint_appl_extra.bib}

\end{document}